\documentclass[twocolumn,superscriptaddress,amsfonts,nopacs,amsmath,amssymb]{revtex4}
\pdfoutput=1
\usepackage{ucs}
\usepackage{graphicx}
\usepackage{graphics}
\usepackage{amsmath}
\usepackage{amssymb}
\usepackage{color}
\usepackage{dcolumn}
\usepackage{epsfig}
\usepackage{bm}
\usepackage{braket}
\usepackage{lscape}

\begin{document}


\title{Anomalous orbital moment in the ferromagnetic phase of the Sr$_{4}$Ru$_{3}$O$_{10}$}
\author{Filomena Forte} 
\affiliation{CNR-SPIN, Via G. Paolo II, 132 I-84084 Fisciano, Italy}
\affiliation{Dipartimento di Fisica ``E.R. Caianiello'', Universit\`a degli Studi di Salerno, Via G. Paolo II, 132 I-84084 Fisciano, Italy}
\author{Lucia Capogna} 
\affiliation{CNR-IOM OGG, 6 rue J. Horowitz, F-38042 Grenoble, France} 
\author{Veronica Granata} 
\affiliation{Dipartimento di Fisica ``E.R. Caianiello'', Universit\`a degli Studi di Salerno, Via G. Paolo II, 132 I-84084 Fisciano, Italy}
%
\author{Rosalba Fittipaldi} 
\affiliation{CNR-SPIN, Via G. Paolo II, 132 I-84084 Fisciano, Italy} 
\affiliation{Dipartimento di Fisica ``E.R. Caianiello'', Universit\`a degli Studi di Salerno, Via G. Paolo II, 132 I-84084 Fisciano, Italy}
\author{Antonio Vecchione} 
\affiliation{CNR-SPIN, Via G. Paolo II, 132 I-84084 Fisciano, Italy} 
\affiliation{Dipartimento di Fisica ``E.R. Caianiello'', Universit\`a degli Studi di Salerno, Via G. Paolo II, 132 I-84084 Fisciano, Italy}
\author{Mario Cuoco} 
\affiliation{CNR-SPIN, Via G. Paolo II, 132 I-84084 Fisciano, Italy} 
\affiliation{Dipartimento di Fisica ``E.R. Caianiello'', Universit\`a degli Studi di Salerno, Via G. Paolo II, 132 I-84084 Fisciano, Italy}
%
%
%
\date{\today}

\begin{abstract}
The coupling of spin and orbital degrees of freedom in the trilayer Sr$_{4}$Ru$_{3}$O$_{10}$ sets a long-standing puzzle, due to the peculiar anisotropic coexistence of out-of-plane ferromagnetism and in-plane metamagnetism. 
Recently, the induced magnetic structure by in-plane applied fields has been investigated by means of spin-polarized neutron diffraction, which allowed to extract a substantial orbital component of the magnetic densities at Ru sites. It has been argued that the latter is at the origin 
of the evident layer dependent magnetic anisotropy, where the inner layers carry larger magnetic moments than the outer ones. We present a spin-polarized neutron diffraction study in order to characterize the nature of the ferromagnetic state of Sr$_{4}$Ru$_{3}$O$_{10}$, in the presence of a magnetic field applied along the $c$-axis. The components of the magnetic densities at the Ru sites reveal a vanishing contribution of the orbital magnetic moment which is unexpected for a material system where orbital and spin degeneracy are lifted by spin-orbit coupling and ferromagnetism.
We employ a model that includes the Coulomb interaction and spin-orbit coupling at the Ru site to address the origin of the suppression of the orbital magnetic moment. The emerging scenario is that of non-local orbital degrees of freedom playing a significant role in the ferromagnetic phase, with the Coulomb interaction that is crucial to make anti-aligned orbital moments at short distance resulting in a ground state with vanishing local orbital moments. 
 
\end{abstract}

\maketitle

\section{Introduction}
The ruthenium perovskites belonging to the Ruddlesden-Popper (RP) series A$_{n+1}$Ru$_n$O$_{3n+1}$ represent a fortunate nature's success to engineer layered materials that can change drastically their electronic and magnetic properties as a function of the number $n$ of RuO$_2$ layers within the unit cell.~\cite{Malvestuto2011} In those compounds, the extended nature of 4$d$ orbitals of the ruthenium ions leads to comparable energies for competing interactions, i.e. crystal field, Hund's interactions, spin-orbit coupling, $p-d$ hybridation and electron-lattice coupling. Moreover, it renders the physical properties highly dependent on the dimensionality ($n$) and susceptible to perturbations such as applied magnetic fields and pressure, without the need for chemical doping~\cite{Qi2012,Cao2016,Lei2014,Forte2010}.
\\
Recently, the key role played by the orbital physics as it concerns the electronic and magnetic properties of layered ruthenates has been invoked for several Ca and Sr based RP compounds. In such systems, the orbital degree of freedom is typically active and has a complex interplay with charge, spin, and lattice degrees of freedom, resulting in a wide range of unique properties including spin-triplet superconductivity in Sr$_2$RuO$_4$,~\cite{Mackenzie2003,Bergemann2003} band-dependent Mott metal-insulator transition,~\cite{Sutter2017} and orbital ordering in Ca$_2$RuO$_4$,~\cite{Das2018,Porter2018} metamagnetism and correlated effects in Sr$_3$Ru$_2$O$_7$.~\cite{Perry2001,Borzi2007}
\\
Situated between $n=2$ and $n=\infty$, Sr$_4$Ru$_3$O$_{10}$ is the $n=3$ member of the Sr- based RP series with triple layers of corner shared RuO$_6$ octahedra separated by SrO rock-salt double layers.~\cite{Malvestuto2013} It displays complex phenomena ranging from tunneling magnetoresistance and low frequency quantum oscillations to switching behavior.
The most intriguing feature, however, is a borderline magnetism: while along the $c$-axis (perpendicular to the Ru-O layers) Sr$_4$Ru$_3$O$_{10}$ shows ferromagnetism with a saturation moment of 1.13 $\mu_B$/Ru and a Curie temperature T$_c$ at 105 K, for the field in the $ab$-plane it exhibits a sharp peak in the magnetization at T$^*$= 50 K and a first-order metamagnetic transition.
The coexistence of the interlayer ferromagnetism and the intralayer metamagnetism, i.e., the anisotropy in the field response, is not typically encountered in magnetic materials and it may arise from a peculiar electronic state with two-dimensional Van Hove singularity close to the Fermi level in conjunction with a distinct coupling of the spins to the orbital states and lattice. Another important ingredient emerging in the Sr$_4$Ru$_3$O$_{10}$ metamagnetism is provided by the magnetoelastic coupling.~\cite{Granata2013,Schottenhamel2016} 
In particular, a direct evidence of a strong spin-lattice interaction has been obtained by means of neutron scattering demonstrating that significant structural changes occur concomitantly with the metamagnetic transition.~\cite{Granata2013}\\
Recently, it has been proposed that a layer-dependent magnetic state may be allowed due to the interplay between octahedral distortions, spin-orbit, and Coulomb interactions.~\cite{Granata2016} In that experiment, a polarized neutron scattering study has been performed in order to analyze the spin and orbital spatial components of the induced magnetization density  $M(\textbf{r})$ with a magnetic field applied in the $ab$-plane and in the metamagnetic regime (B $>$ 2 Tesla). It was found that there exists an order relation between spin and orbital moments and their amplitudes in the unit cell, since they are strongly linked to the layers where the electrons are located. In the specific, the inner ruthenium ions in the triple layer have larger spin and orbital magnetic moments than the outer ones. Remarkably, the inner-outer correspondence is robust with respect to temperature variations, since it persists even above T$_c$, thus indicating that these features are intrinsic to the high-field magnetic state.\\
We present here the outcome of a a polarized neutron scattering study on the same high quality single crystal of Sr$_4$Ru$_3$O$_{10}$, for the case where the magnetization density is in the ferromagnetic regime i.e. with a strong magnetic field applied along the easy axis ($c$-axis). This study is motivated by the need to clarify the nature of the ferromagnetic phase and in particular the role of the orbital magnetic moment in the inter/intra layer magnetism of this system. 
The refinement of our neutron scattering data reveals a vanishing contribution of the orbital component of the magnetic density at Ru sites within the unit cell, which is contrary to the case of an in-plane applied field, where a substantial orbital angular momentum was measured.
We employ an effective correlated model, that includes the coupling between the spin-orbital degrees of freedom at inequivalent Ru sites, to interpret the orbital quenching of the high-field magnetic state. We ascribe the origin of the vanishing orbital component along the $c$ direction to the development of robust short-range antiferro-orbital correlations, which arise together with the onset of the fully polarized spin state. Remarkably, our analysis demonstrates the major role played by the orbital angular momentum in determining the anisotropy of the magnetic response for field applied along the $c$ axis with respect to the case of af in-plane applied field. Due to the peculiar competition between electron correlations, spin-orbit effects and the layer-dependent crystal-field (CF) splitting set by the octahedral environment, the orbital component of the magnetic moment is ubiquitous suppressed within each layer of the unit cell when a longitudinal field is applied, while it is substantial and at the origin of the inequivalent intra-layer and inter-layer magnetic response in the case of an in-plane field.\\
The paper is organized as follows. In Sec. II, we introduce the experimental setup and the polarized beam approach. In Sec. III, we present the experimental results, while Sec. IV is devoted to their interpretation and description of a theoretical model which is able to explain the quenching of the orbital angular momentum in the high field magnetic state. In Sec. V , we provide the concluding remarks.

\begin{table*}[t]
	\centering
	\caption {Total and orbital contributions to the magnetic densities at the ruthenium and oxygen sites, at the three relevant temperatures of the system, as refined with the program Fullprof.} \label{Tab1}
	\resizebox{1\textwidth}{!}{
		\begin{tabular} {c c c c c c c}
			\hline
			\hline
			&  & \multicolumn{2}{c}{$\phantom{....................}$}  & \multicolumn{2}{c}{ $\phantom{....................}$}\\
			\multicolumn{1}{c}{\textbf{ \large B // c $\quad$}} & \multicolumn{2 }{c}{\textbf{ \large 2 K}} & \multicolumn{2}{c}{\textbf{$\qquad\qquad$ \large 50 K}} & \multicolumn{2}{c}{\textbf{$\qquad\qquad$ \large 115 K}}\\ 
			&  & \multicolumn{2}{c}{$\phantom{ ...................................}$}  & \multicolumn{2}{c}{ $\phantom{....................}$}\\
			\hline
			\rule[0mm]{0mm}{6mm}
			\large M$(\mu_B)$ $\quad$ & $\quad$ \large M$_{tot}$ & $\quad$ \large M$_{orb}$    & $\quad$ $\qquad$  \large M$_{tot}$ & $\quad$ 
			\large M$_{orb}$ & $\quad$$\qquad$  \large M$_{tot}$ & $\quad$$\quad$  \large M$_{orb}$\\
			\rule[0mm]{0mm}{6mm}
			\large Ru$_{in}$  $\quad$ & $\quad$  \large 1.75(7)  & $\quad$ \large 0.1(2)   &  $\quad$$\qquad$  \large     1.5(7) &  $\quad$ \large 0.1(2) \quad  &   $\quad$$\qquad$  \large    0.95(7) &  $\quad$  $\quad$\large 0.1(2)\\  
			\rule[0mm]{0mm}{6mm}
			\large Ru$_{out}$ $\quad$   & $\quad$  \large 1.10(6)  & $\quad$  \large -0.1(1)   &   $\quad$$\qquad$  \large  1.0(1) &  $\quad$   \large -0.1(1)   &  $\quad$$\qquad$    \large  0.61(6) &  $\quad$  $\quad$ \large -0.1(1)    \\ 
			\rule[0mm]{0mm}{6mm}
			\large O$_{bas}$ $\quad$   & $\quad$ \large 0.15(1)    & $\quad$&     $\quad$$\qquad$  \large 0.13(1)   &$\quad$ & $\quad$$\qquad$ \large  0.07(1) \\   
			\rule[0mm]{0mm}{6mm}
			\large O$_{ap}$  $\quad$ &   $\quad$  \large 0.05(2)   &$\quad$ &       $\quad$$\qquad$  \large 0.04(2)   &$\quad$ &     $\quad$$\qquad$  \large 0.03(1) \\
			\\
			\hline
			\rule[0mm]{0mm}{7mm}
			\large $\chi^2$  $\quad$ &   $\quad$  \large 1.74   &$\quad$ &       $\quad$$\qquad$  \large 2.60   &$\quad$ &     $\quad$$\qquad$  \large 1.81 \\
			\\
			\hline
			\hline
		\end{tabular}
	}
\end{table*}

\section{Experimental methodology}

Single crystals of Sr$_{4}$Ru$_{3}$O$_{10}$ were grown in an image
furnace as described elsewhere.~\cite{Fittipaldi2007} The samples
were cut into small rectangular slices with an average size of
4x4x0.2 $mm^{3}$. Similar samples were used in our previous studies.~\cite{Granata2013,Granata2016} X-ray
diffraction, energy and wavelength dispersive spectroscopy as well
as neutron Laue diffraction have been used to fully characterize
the structure, quality and purity of the crystals. Magnetizations measurements on crystals from the same batch identified the ferromagnetic transition at T$_c\cong105$ K and a
metamagnetic transition at the temperature 
T$^*\cong50$ K, when a magnetic field is applied in the {\itshape a\itshape b}-plane. The metamagnetic transition appears for magnetic field up to 2 Tesla.~\cite{Carleschi2014}\\
The experiments were carried out on the D3 neutron diffractometers at the Institut Laue Langevin in Grenoble. 
A magnetic field of 9 Tesla was applied on cooling along the $[00l]$ direction, the $c$-axis,  which is the magnetic easy direction for the crystal. Under these conditions, one could be confident that the magnetic moments were completely aligned in the vertical axis, so that the expression for the flipping ratio in its simplified form is valid~\cite{Squires}.  
A wavelength of about 0.825 $\AA$ was used to measure the scattering in the $[h00/0k0]$ plane with a tilting option for the detector in order to acquire the intensity of many $(hkl)$ reflections with non vanishing $l$ index.  Since the crystal cell is quite extended along $z$, $c\sim 28$ $\AA$, values of $l$ up to 13 could be reached. \\
A radiofrequency coil inserted between the monochromator and the sample was used to flip the spin state of the incident neutrons so that the intensity of about 65 independent reflections could be measured at three different temperatures: 2, 50 and 120 K in both the up and down spin polarization, with a degree of polarization of 0.94. A 0.5-$mm$ erbium filter allowed to reduce higher-order contamination in the incident beam. A Heussler monochromator was used. \\
The cell parameters together with the exinction coefficients which are required to calculate the structure factors were previously determined at 115 K on D9 diffractometers within the primitive $Pbam$ space group. An alternative description of the crystal as a face-centered structure with some degree of oxygen disorder along the $c$-axis was also tested using the space group 64 in the $Acam$ choice of axis. Since both crystal structures lead to the same results in terms of resulting magnetization densities, in the final analysis we made use of the $Pbam$ space group, which is well documented in the literature.~\cite{Crawford02}\\
The measured flipping ratios were then loaded on the program Fullprof~\cite{fullprof} to refine the magnetic moments localized on the atoms of ruthenium but also on the oxygens. The spherical approximation for the electron density  proved to be good enough to model the density of magnetization.~\cite{Carvajal1993}

\section{Spin polarized neutron diffraction study}
\begin{figure}[!h]
	\begin{center}
\includegraphics[width=8cm,height=12cm,angle=0]{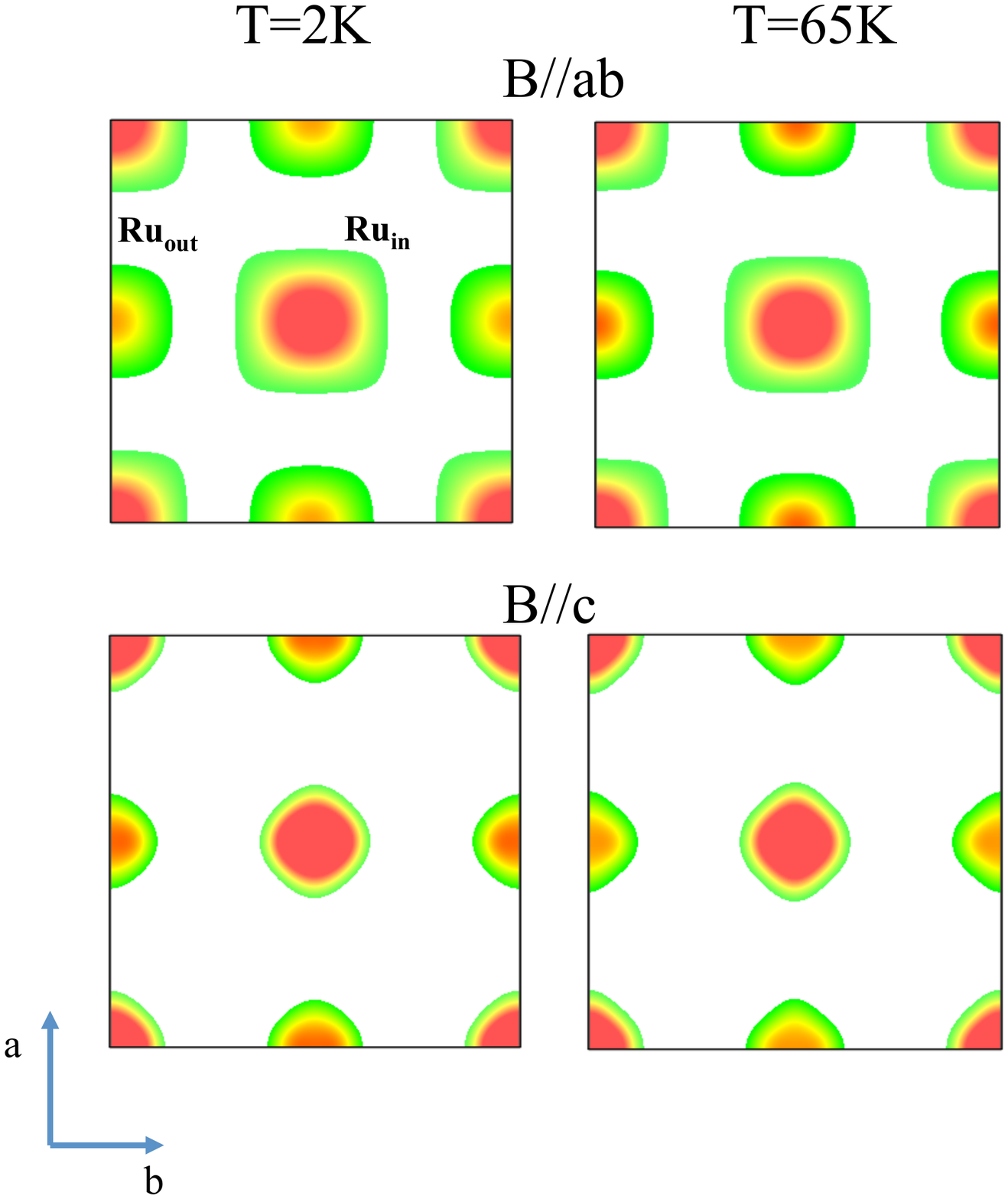}
		\caption {\footnotesize Sections of the magnetization density in the $ab$ plane of Sr$_4$Ru$_3$O$_{10}$ calculated directly with the maximum entropy method. The two maps on the left are for a sample polarization in the $ab$ plane, whilst the right ones are for a field applied along the $c$-axis.} \label{Fig.1}
	\end{center}
\end{figure}

In Table~\ref{Tab1}, the values of magnetic densities at the ruthenium sites located in the inner and outer octahedra of the triple-layer unit cell are reported, together with those of the oxygens, for the three different temperature values of 2, 50 and 120 K. As previously 
reported,~\cite{Granata2016} the contribution on the oxygens in this kind of systems is not negligible. Several important observations can be extracted from the results in Table~\ref{Tab1}.
First of all, we point out that the orbital component of the magnetic moment appears to be vanishing in the present field configuration. Indeed, its value is undetectable within the experimental error, at each of the temperature which have been considered. This crucial result points out a major difference with respect the to what was previously found in the case of an in-plane applied field, where a substantial orbital contribution to the magnetic densities was detected at Ru sites.~\cite{Granata2016} 
The discussion of the microscopic mechanism leading to this anisotropic orbital quenching in the high field magnetic states will be presented in the next section.\\
Another important observation is concerned with the layer dependence of the magnetic (spin) density, which is graphically depicted in the maps shown in Figure~\ref{Fig.1}. Those magnetization density maps were calculated in a direct way with the maximum entropy method 
~\cite{Squires} implemented in the program Dysnomia 
 which is now available in the Fullprofsuite. The most likely spin distribution is the one that maximizes the entropy among those which are compatible with the observed magnetic structure factors. The latter were refined with the program Fullprof.\\
Our results show that the inner Ru ions carry a larger spin momentum with respect to the outer Ru ions, and that this order relation is robust against temperature variation. In Figure~\ref{Fig.1}, the present results ($B//c$) are compared to what previously reported for $B//ab$.~\cite{Granata2016} The most noticeable effect of the field polarization along the $c$-axis is to invert the trend of the magnetization density on the outer ruthenium ions when the temperature is swept across the metamagnetic transition at around 50 K. Indeed, with the field in the 
$ab$-plane, the outer ruthenium ions are more intensely magnetized in the vicinity of T* than at low temperature. On the contrary,  the opposite behaviour occurs with the field along $c$, in the latter case the outer ruthenium ions  appearing to be more intensely magnetized at 2 K than at T*.\\
The inequivalence of the inner and outer ruthenium sites has crystallographic origins as well as physical reasons. At zero field, the outer octahedra being slightly elongated contrary to the inner ones which are regular octahedra.~\cite{Crawford02} In addition, the inner octahedra are more rotated than the external ones,  with an angle of rotation above the critical angle for ferromagnetism.~\cite{Singh2001} As a consequence, the Ru$_{in}$ are supposed to be more prone to ferromagnetism than the Ru$_{out}$. This is reflected in the value of ferromagnetic moment along the $c$-axis on the two sites, which has been measured with neutron scattering at low temperature (2 K) yielding 1.59 $\mu B$ on the inner ruthenium ions and 0.92 $\mu B$ on the outer ones.~\cite{Granata2013} In the next section, we focus on the role of the orbital component of the magnetic moment and discuss a mechanism which leads to the anisotropic suppression, which depends on the orientation of the applied field.\\


\section{Modelling the ferromagnetic state with vanishing orbital magnetic moment}

In this section, we propose a physical scenario which is able to account for the occurrence of a vanishing orbital magnetic moment in the high-field ferromagnetic phase of Sr$_4$Ru$_3$O$_{10}$. The analysis is performed by focusing on the orbital character of the spin-polarized ground-state described by an effective microscopic model that we solve on a cluster in order to include on equal-footing all the interacting electronic degrees of freedom for the $d$-states at the Ru site. 
The physical context is set by our polarized neutron diffraction study, which reveals a layer independent quenching of the orbital component of the magnetic density, thus manifesting both at the Ru ions belonging to the central and the outer RuO planes of the unit cell, assuming an external field is applied along the $c$-axis. This result has a completely opposite outcome and trend if compared to a previous analogous study, which has instead demonstrated the occurrence of a substantial orbital component in the magnetic density at the Ru sites, when a field of equal strength is applied along the $ab$-plane in the metamagnetic phase.\\
Our aim is to provide a microscopic scenario to account for the realization of a spin-polarized phase with almost zero orbital moment. 
On a general ground, the magnetic moment carried by electrons in solid state materials has two components: the one arising from its spin and the one
originating from its orbital character. The local spin magnetic moment typically 
emerges as a consequence of the Coulomb interaction and, in partiicular, the Hund's coupling is a key player at work in the majority of the
magnetic solids. On the other hand, concerning the formation of the
orbital magnetic moment, it is due to the spin-orbit coupling, which lifts the
quenching of the orbital moment in a magnetic solid. 
Hence, in a ferromagnetic configuration, where the spin degeneracy is lifted, the spin-orbit coupling (SOC), $H_{SOC}=\lambda {\bf L} \cdot {\bf S}$, is expected to yield an orbital moment ${\bf L}$ tied to the spin moment $({\bf S})$, with $\lambda$ being the strength of the SOC.
We would then have a local orbital moment at each Ru site along the $c$-axis in the spin-orbit coupled ferromagnetic state of Sr$_4$Ru$_3$O$_{10}$, hence another mechanism has to be invoked to account for its suppression.
Starting from this picture, we demonstrate that the formation of distinct non-local orbital correlations can be the driving mechanism leading to the suppression of the orbital moment. More specifically, a correlated ferromagnetic state with isotropic short-range antiferro-type orbital configurations, i.e. orbital moments aligning in a way to be antiparallel on neighboring sites, 
can yield a quantum configuration where the average on-site orbital moment is suppressed.
In order to proceed further, we introduce a minimal model which is able to capture the competition between the local and non-local orbital physics by studying an effective two-sites Ru cluster and including all the relevant microscopic ingredients, i.e. Coulomb interaction, spin-orbit coupling, and crystalline field potentials associated with the RuO$_6$ octahedral distortions. 

Hence, the examined microscopic 
model Hamiltonian with two inequivalent atoms Ru$_{1}$ and Ru$_{2}$ is expressed as: 
\begin{eqnarray}
H=H_{kin}+H_{el-el}+H_{cf}+H_{so}+H_{z} \,. \label{totham}
\end{eqnarray}
The first term in Eq. \ref{totham} is the kinetic operator between the $t_{2g}$ orbitals on different Ru sites
\begin{eqnarray}
H_{kin}=-t \sum_{ij,\sigma} ( d^{\dagger}_{{\bf i} \alpha \sigma}
d_{{\bf j} \alpha \sigma} +h.c.)
\end{eqnarray}
$d^{\dagger}_{i \alpha \sigma}$ being the creation operator for an
electron with spin $\sigma$ at the $i$ site in the $\alpha$
orbital. 
The second term is the local Coulomb interaction
between $t_{2g}$ electrons:
\[
\nonumber H_{el-el} =  U \sum_{i\alpha} n_{{\bf i} \alpha
\uparrow} n_{{\bf i} \alpha \downarrow} - 2 J_{H} \sum_{i \alpha
\beta} {\bf S}_{{\bf i} \alpha}\cdot {\bf S}_{{\bf i} \beta}+
\]
\begin{equation}
(U'-\frac{J_{H}}{2}) \sum_{i\alpha\neq\beta} n_{{\bf i} \alpha}
n_{{\bf i} \beta}+J' \sum_{i\alpha \beta} d^{\dagger}_{{\bf i}
\alpha \uparrow} d^{\dagger}_{{\bf i} \alpha \downarrow} d_{{\bf
i} \beta \uparrow} d_{{\bf i} \beta \downarrow}
\end{equation}
\noindent where $n_{{\bf i} \alpha \sigma}$, ${\bf S}_{{\bf i}
\alpha}$ are the on site charge for spin $\sigma$ and the spin
operators for the $\alpha$ orbital, respectively. $U$ ($U'$) is
the intra (inter)- orbital Coulomb repulsion, $J_{H}$ is the Hund
coupling, and $J'$ the pair hopping term. Due to the invariance
for rotations in the orbital space, the following relations hold:
$U=U'+2 J_{H}$, $J'=J_{H}$. \\
The $H_{cf}$ part of the Hamiltonian $H$ is the crystalline field potential,
controlling the symmetry lowering from cubic to tetragonal one:
\begin{eqnarray}
H_{cf}= \sum_i \Delta_i [n_{{\bf i}xy}-\frac{1}{2}(n_{{\bf i}xz}+n_{{\bf i}yz})]
\,
\end{eqnarray}
	
Positive (negative) values of $\Delta$ are related to elongated
(flat) RuO$_6$ octahedron along the $c$-axis and favor the
occupation in the $d_{xz,yz}$ ($d_{xy})$ sector, respectively. 
For the present investigation, we start by considering as a reference the CF configuration at zero field, and assume  $\Delta_1$ to be negative, to simulate the Ru$_{in}$, and $\Delta_2$ to be positive, as to simulate the Ru$_{out}$. In second instance, we will also vary the CF parameters, 
in order to address possible magneto-elastic effects driven by the applied field and to assess the robustness of the obtained effects with respect to the octahedral distortions. Since we deal with 4$d$ oxides, it is also important to include the atomic spin-orbit
coupling
\begin{equation}
H_{so}=\lambda \sum_i {\bf L_{i}}\cdot {\bf S_{i}} \,
\end{equation}
\noindent where $\textbf{L}_i$ is the on-site total orbital momentum  and $\textbf{S}_i$ the total spin of the $t_{2g}$ states. \\
Finally, $H_{z}$ in Eq. \ref{totham} describes the Zeeman coupling of the local angular momenta to a magnetic field $B$ applied along the $c$ symmetry direction, expressed in unit of the 
Bohr magneton:
\begin{equation}
H_{z}=\sum_{i} (\textbf{L}_i+2\textbf{S}_i)\cdot \textbf{B}_{z} \,.
\end{equation}
We follow LDA predictions~\cite{Malvestuto2011} and assign as a reference in the calculations the values: $t=0.4eV$ and $\Delta_{1}/t=-0.3$, $\Delta_{2}/t=0.225$ to the CF parameters of 
the central-flat and outer-elongated octahedra, respectively. Subsequently, we will also vary $\Delta_{2}/t$ from positive to negative values, in order to explore different regimes of octahedral distortions. Concerning the Coulomb interactions, we consider the ratio $U/J_H=5.0$ and analyze the regime of intermediate 
electronic correlations set by $U/t=5.0$.~\cite{Gorelov2010} 
Concerning the SOC term we assume an amplitude $\lambda/t=0.14$, in the range of values that is expected to hold for Sr-based ruthenates.~\cite{Haverkort2008} Hence, to fully characterize the spin and orbital character of the ground-state, we determine the average expectation values of the spin and orbital moments as well as the nearest-neighbor orbital correlations computing the ground-state configuration of the model Hamiltonian in Eq. \ref{totham} by means of exact-diagonalization, assuming a cluster made by two inequivalent ruthenium sites Ru$_1$ and Ru$_2$. \\

\begin{figure}[t!]
\begin{center}
\includegraphics[width=8.2cm]{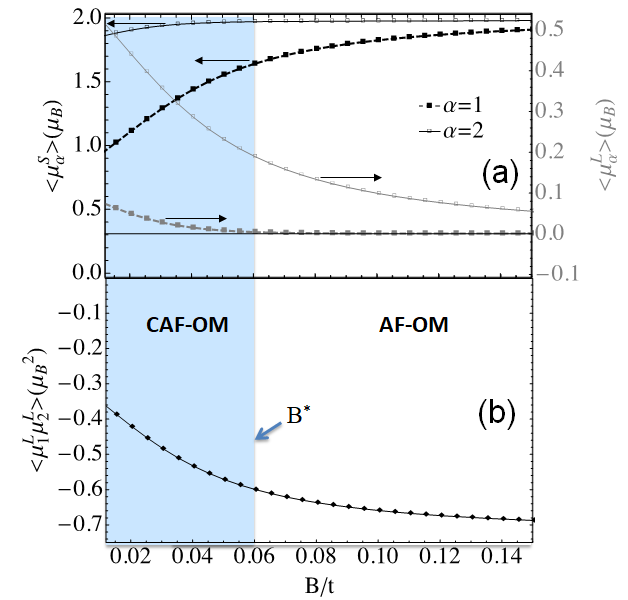} 
\end{center}
\caption{(Color online) Evolution of (a) the spin and orbital components  of the Ru magnetic density as well as (b) of the Ru-Ru non-local orbital moment correlations projected along the $c$-axis. $B$ is the amplitude of the applied Zeeman field oriented along the $c$-axis. We assume that $t=0.4eV$, $\Delta_{1}/t=-0.3$, $\Delta_{2}/t=0.225$, $U/t=5.0$, $J_H/U=0.2$. AF-OM (CAF-OM) stands for a ferromagnetic ground-state with anti-aligned orbital moments and inequivalent resulting orbital polarization with an averaged amplitude $\mu^{L}_{av}=(1/2)(\langle \mu^{L}_{1}\rangle+\langle \mu^{L}_{2}\rangle)$ that is smaller (larger) than 0.1. $B^{*}$ indicates the amplitude of the Zeeman field above (below) which the ground state is spin polarized with a small (large) averaged orbital moment than a given reference that is set at 0.1.}\label{fig:theo1}
\end{figure}

We start by considering the evolution of the ferromagnetic state as a function of an applied Zeeman field along the $c$-axis to assess the relation between the amplitude of the local spin and orbital magnetic moment and the nearest-neighbor orbital pattern. 
In Fig. \ref{fig:theo1}(a) we present the field dependence of the $c$-axis projection of the local spin, $m_S$, and orbital, $m_L$, components of the magnetic density evaluated at the Ru$_{1}$ and Ru$_2$ sites. Together with the on-site spin and orbital moment, we track the evolution of the non-local orbital correlations between the orbital moments at the two Ru sites (Fig. \ref{fig:theo1}(b)). The analysis is performed for a representative distortive state of the octahedra at the Ru$_1$ and Ru$_2$ sites being in a flat and elongated configuration, respectively. 
There are various remarkable aspects of the correlated ferromagnetic state that the presented investigation unveils.
Firstly, as one would expect, we observe that inequivalent octahedral distortions generally lead to different spin and orbital magnetic moments at the corresponding Ru sites. Furthermore, we have to remind that the orbital degree of freedom in the $d^4$ Ru configuration is set by the position of the doubly occupancy within the three t$_{2g}$ orbital states. Then, since flat (elongated) octahedra tend to favor an orbital occupancy with the doublon placed in the $xy$ ($\{xz,yz\}$) orbital sector, one has that the spin-orbit lifting of the orbital degeneracy leads to an orbital magnetic moment which is preferably oriented in the $ab$ plane ($c$-axis), respectively. 
The major consequence of the interplay between octahedral distortions and spin-orbit coupling is that the magnetic anisotropy would make an easy spin-polarization in the $ab$ plane or along the $c$-axis depending on whether the octahedra are flat or elongated. Such qualitative trend is confirmed by the outcome of the theoretical analysis (Fig. \ref{fig:theo1}(a)) and also is consistent with the larger amplitude of the spin and orbital moments at the Ru$_1$ site when compared with Ru$_2$.    

As we can see in Fig. \ref{fig:theo1}, the increase of the applied magnetic field can drive the crossover between two distinct ferromagnetic configurations. Indeed, at low field the ground-state, labelled as CAF-OM, exhibits a significant local spin moment and an orbital component, whose amplitude is comparable to the spin polarization (see Fig. \ref{fig:theo1}). On the other hand, the Ru-Ru orbital correlations are negative and thus indicate that the orbital moments are anti-aligned. The amplitude of the Ru-Ru orbital-moments correlator, however, is not maximal in the CAF-OM and thus it shows a sort of canting in the orbital configuration. A further growth of the applied field tends to suppress the averaged orbital moment over the two Ru sites. Indeed, as reported in Fig. \ref{fig:theo1}(b), one  can single out an effective amplitude of the Zeeman field, $B^{*}$, that separates the CAF-OM state from the AF-OM one Characterized by a reduced averaged orbital moment $\mu^{L}_{av}=(1/2)(\langle \mu^{L}_{1}\rangle+\langle \mu^{L}_{2}\rangle)$, lower than the reference amplitude of $\mu^{L}_{av}=0.1$. 

\begin{figure}[t!]
\begin{center}
\includegraphics[width=8.2cm]{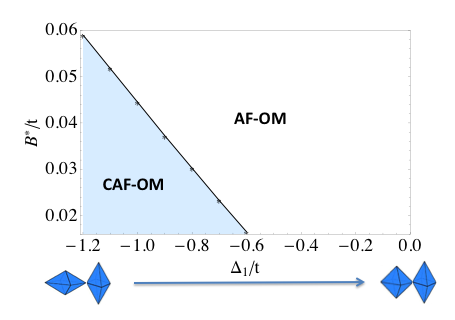} 
\end{center}
\caption{(Color online) Evolution of the AF-OM and CAF-OM ferromagnetic phases as a function of the $c$-axis Zeeman field and of the octahedral distortions at the Ru$_1$ site ($\Delta_1$), for a given amplitude of the crystal field potential (elongated configuration) at Ru$_2$. $B$ is the amplitude of the applied magnetic field. We assume that $t=0.4eV$, $\Delta_{1}/t=-0.3$, $U/t=5.0$, $J_H/U=0.2$. AF-OM (CAF-OM) stand for ground-state with anti-aligned orbital moments and inequivalent resulting orbital polarization corresponding to an averaged amplitude $\mu^{L}_{av}=(1/2)(\langle \mu^{L}_{1}\rangle+\langle \mu^{L}_{2}\rangle)$ that is smaller (larger) than 0.1. $B^{*}$ indicates the amplitude of the Zeeman field above (below) which the ground state is spin polarized with a small (large) averaged orbital moment than a given reference that is set at 0.1.}\label{fig:theo2}
\end{figure}

To assess the relation between the crossover field amplitude $B^{*}$ and the character of the octahedral distortions, we explore the impact on the ground state of the variation of the crystal field potential. In particular, we keep in mind that recent studies revealed that distinct structural mechanisms are associated with the $c$ axis magnetization, specifically an induced extension of the $c$ axis, which may result in an elongation of the apical RuO bonds and a contraction of the in-plane RuO bonds.\cite{Schottenhamel2016} In our calculations, we mimic this trend by lowering the CF term at the Ru$_1$ site, by keeping fixed the configuration at the Ru$_2$ site. 
In Fig. \ref{fig:theo2}, we report the evolution of the CAF-OM and AF-OM phases as a function of $\Delta_1$ by moving from the zero field compressed octahedral state ($\Delta_1/t\sim$ -1.2) towards the symmetric configuration ($\Delta_1/t$=0). The outcome sets distinctive marks of the spin-polarized ferromagnetic states with respect to the local octahedral distortions. The two ferromagnetic phases are indeed separated only in the regime of strong octahedral inequivalence at the neighboring Ru sites, where the suppressed orbital magnetic moment AF-OM phase is obtained only upon the application of a Zeeman field above a given threshold $B^{*}$. On the other hand, when the flattening at The Ru$_1$ site is released by simulating a weakly distorted octahedron ($\Delta_1/t$ $\sim$ 0) , the ferromagnetic state with a vanishing orbital moment is settled at low fields and holds at any amplitude of the applied magnetic field.\\

Finally, it is worth pointing out that the main objective of the performed computation is to unveil the mechanisms that can account for the observed orbital quenching. Given the clusters size of the quantum simulation and the many competing energy scales in the problem, a tight quantitative correspondence between the theoretical outcome and the experimental data is naturally beyond the scope of the effective model. Still, our analysis is more suitable to single out the range of the microscopic parameters where a reasonable qualitative and quantitative agreement on the trend and evolution of the physical  observables can be achieved.
From this point of view, the specific choice of the Coulomb interaction parameters, $U$ and $J_H$, is not crucial. However, the disentangling of spin and orbital degrees of freedom by the applied magnetic field along $c$, which drives magnetic ordering coexisting with short-range antiferro-orbital correlations can only be obtained in a correlated picture.\\
\newline

\section{Conclusions}

We used polarized neutron scattering diffraction to determine the spin and orbital character of the magnetic state of Sr$_4$Ru$_3$O$_{10}$, in the high-field ferromagnetic phase with spin-moments aligned along the $c$-axis. We discussed the microscopic mechanisms which are able to account for the suppression of the orbital moment, as due to the formation of robust antiferro-orbital correlations in the highly spin polarized phase. These results are in striking contrast with what was obtained for a field applied in plane, but they can be reconciled within the same correlated picture where local Coulomb interaction, structural changes of the octahedra, SOC are strongly competing. Our findings unveil two distinct ferromagnetic states with a completely distinct orbital moment configuration, which are tuned by an applied magnetic field. The effective magnetic field to access the phase with a vanishing orbital moment is also strongly tied to the character of the octahedral distortions affecting the Ru orbitals, being lower in the case where the octahedra are more uniform and less compressed along the $c$ axis. We point out that our comparative studies of the field induced magnetic phases of Sr$_4$Ru$_3$O$_{10}$, for longitudinal and in-plane applied fields, highlight the role of the orbital component in setting the anisotropic magnetic response. In particular, we provide a physical scenario where the orbital component is almost fully quenched when the field is applied along the $c$ axis while it is substantial and at the origin of the layer-dependent magnetic response, in the case of an in-plane applied field.

 \begin{acknowledgments}
 We acknowledge insightful discussions and valuable 
 support during the experiments by Anne Stunault, Oscar Fabelo, Alberto José Rodriguez
 Velamazan and Arsène Goukassov.
 \end{acknowledgments}

%





\end{document}